\documentclass[fleqn,usenatbib]{mnras}
\usepackage{newtxtext,newtxmath}
\usepackage[T1]{fontenc}
\usepackage{ae,aecompl}

\usepackage{graphicx}	% Including figure files
\usepackage{amsmath}	% Advanced maths commands
\usepackage{amssymb}	% Extra maths symbols
\usepackage{pifont}
\usepackage{siunitx}
\DeclareSIUnit\year{yr}
\usepackage[dvipsnames]{xcolor}

\title[A Geometric Probe of Cosmology: I. Lensing \& Reverberation Mapping]{A Geometric Probe of Cosmology: I. Gravitational Lensing Time Delays and Quasar Reverberation Mapping}

\author[Ng \& Lewis]{
Angela L.H. Ng\thanks{E-mail: angela.ng@sydney.edu.au} and
Geraint F. Lewis
\\
Sydney Institute for Astronomy, School of Physics, A28, The University of Sydney, NSW, 2006, Australia\\\
}

\date{Accepted XXX. Received YYY; in original form ZZZ}

\pubyear{2018}

\begin{document}
\label{firstpage}
\pagerange{\pageref{firstpage}--\pageref{lastpage}}
\maketitle

\begin{abstract}
We present a novel, purely geometric probe of cosmology based on measurements of differential time delays between images of strongly lensed quasars due to finite source effects.
Our approach is solely dependent on cosmology via a ratio of angular diameter distances, the image separation, and the source size. It thereby entirely avoids the challenges of lens modelling that conventionally limit time delay cosmography, and instead entails the lensed reverberation mapping of the quasar Broad Line Region.
We demonstrate that differential time delays are measurable with short cadence spectroscopic monitoring of lensed quasars, through the timing of kinematically identified features within the broad emission lines.
This provides a geometric determination of an angular diameter distance ratio complementary to standard probes, and as a result is a potentially powerful new method of constraining cosmology.
\end{abstract}


%%%%%%%%%%%%%%%%%%%%%%%%%%%%%%%%%%%%%%%%%%%%%%%%%%

%%%%%%%%%%%%%%%%% BODY OF PAPER %%%%%%%%%%%%%%%%%%

\section{Introduction}

The time delay between multiple images of strongly gravitationally lensed quasars is a cosmological probe \citep{Refsdal1964, BlandfordNarayan1992, Suyu2014, Treu2013}. Such a time delay is a result of the geometric and gravitational differences in the light paths corresponding to each image. If the lensed source is time variable, such as a quasar, the time delay is detectable via photometric monitoring.

The gravitational lensing time delay is therefore a direct physical measurement of cosmological distances, from which we are able to constrain cosmological parameters. Conventional time delay measurements are most sensitive to the Hubble constant $H_0$, to which it is inversely proportional.
Independent tests of cosmological parameters are especially important given the current tension between determinations of $H_0$ at low redshifts from the cosmic ``distance ladder" and from CMB data. This tension is either due to unknown systematics or new physics \citep{PlanckCollaboration2016, DiValentino2016, Alam2017, Riess2018, Riess2019}. It follows that there is increasing interest in time delay cosmography \citep[e.g.][]{Tewes2013, Courbin2018, Birrer2019a}. One such project is $H_0$ Lenses in COSMOGRAIL's Wellspring \citep[H0LiCOW;][]{Suyu2017}, with a current 3.8 per cent precision measurement of $H_0$ \citep{Bonvin2017} as part of the COSmological MOnitoring of GRAvItational Lenses \citep[COSMOGRAIL;][]{Courbin2004} program.

The main caveat with regards to time delay cosmography is that gravitational lens modelling depends implicitly upon the assumed underlying mass distribution. In particular, there is a systematic problem associated with the ``mass sheet degeneracy" where the degeneracy of ill-constrained lens models leads to uncertainties in estimations of $H_0$ \citep{Falco1985, Saha2000}.

In this paper we present a novel geometric test of cosmology that is \textit{independent} of the lensing potential, by considering \textit{differential} time delays \textit{over} images, originating from spatially-separated photometric signals within a strongly lensed quasar. Measuring these differential time delays, in addition to the standard time delay, will give bounds on cosmological parameters that are essentially based on simple geometry. This is in contrast with conventional methods of determining cosmological parameters, such as steps in the ``distance ladder", which are vulnerable to the details of complicated astrophysics (e.g. supernova explosions and structure formation) and to accumulating systematic errors, as each method is used to calibrate the next \citep{Riess2018}.

The paper is organised as follows: we review conventional gravitational time delay measurements in Section \ref{sec:background} and in Section \ref{sec:timedelaydifferences} we introduce the differential time delays across images. We review reverberation mapping in Section \ref{sec:reverbmapping}, and in Section \ref{sec:lensedrm} combine reverberation mapping with gravitational lensing. Section \ref{sec:spectra} describes a method for measuring differential time delays and we discuss in Section \ref{sec:timescales} the relevant timescales and observational prospects. We outline our conclusions and directions for future work in Section \ref{sec:conclusions}.

\section{Gravitational Lensing} \label{sec:gravlensing}
\subsection{Background} \label{sec:background}
If a gravitational lens produces multiple images of a source, the time required for light to reach an observer will be, in general, different for different paths. The time taken for a given path can be found in the standard cosmological context from the null geodesics of the perturbed Friedmann-Lema\^{i}tre-Robertson-Walker (FLRW) metric.

Consider a photometric signal originating from a point source at a position $\vec{\beta}$ on the sky and at line-of-sight physical distance $l_s$ which, if the source were unlensed, an observer would see at time $t=t_i$. As the source is lensed, an observer sees this signal in a given lensed image X at a position $\vec{\theta}_X$ and at time $t_X = t_i + \tau_X$, where $\tau_X$ is the time delay \citep[e.g.][]{Schneider1992, BlandfordNarayan1992}:
\begin{equation}
    \tau_X \equiv \tau \left(\vec{\theta}_X , \vec{\beta}, l_s\right) = \frac{D}{c} ( 1 + z_d ) \left(\frac{1}{2} ( \vec{\theta}_X - \vec{\beta})^2 - \Psi ( \vec{\theta}_X )\right). \label{tauX}
\end{equation}
Here $D= \frac{D_d D_s}{D_{ds}} \propto \frac{1}{H_0}$ is the ``lensing distance", a ratio of angular diameter distances (subscript $d$, $s$, $ds$ denoting the angular diameter distance to the lens, source and between the lens and source, respectively). The lensing distance is thus the factor containing all the cosmological information. We denote the speed of light by $c$, and the redshift of the lensing mass by $z_d$. The two-dimensional vector positions of image X in the lens plane, and the source in the source plane, are given by $\vec{\theta}_X$ and $\vec{\beta}$ respectively; scaled such that their magnitudes are the observed angular positions relative to the observer-lens axis. The dimensionless ``lensing potential" is denoted by $\Psi$.

The first term in Equation \eqref{tauX} is a geometric component, arising from the difference in path length of a lensed versus unlensed photon; and the second term is a potential term accounting for the gravitational time dilation caused by the lensing mass. We note that the time delay $\tau$ is not an observable quantity, since it is the delay relative to an unlensed photon. We can, however, observe time delays between images, e.g. between images A and B:
\begin{align}
\begin{split}
\Delta \tau_{BA} \equiv& \, \tau_B - \tau_A\\
=& \, \frac{D}{c} (1 +z_d) \Big(\frac{1}{2}  \left(\vec{\theta}^2_B - \vec{\theta}^2_A \right) + (\vec{\theta}_A - \vec{\theta}_B) \cdot \vec{\beta} \\
&- \Psi(\vec{\theta}_B) + \Psi (\vec{\theta}_A)  \Big). \label{tauBA}
\end{split}
\end{align}
Measuring time delays between images of strongly lensed quasars is a conventional method employed to test cosmology. However, since Equation \eqref{tauBA} is dependent on the  dimensionless lensing potential $\Psi$, time delay measurements are limited by the assumptions and accuracy of the lens model. Simple gravitational lens systems are rare, and observational constraints on the lens model are limited due to the existence, typically, of only two or four images per system \citep{Schneider2013, Birrer2019a}.

Furthermore, \citet{Falco1985} showed that lens models are in fact degenerate. All observables (such as relative image positions and magnification ratios) are invariant, \textit{except for $H_0 \Delta \tau$}, under a family of transformations of the mass profile of the lens along with a translation of the unobservable source position. Any given set of measurements of image positions and fluxes in a lens system therefore can be consistent with a number of different lens models, and therefore a number of different values of $H_0$ whilst preserving the observed time delay $\Delta \tau$. Uncertainties in the estimated values of cosmological parameters from different gravitational lensing systems, and even in separate analyses of the same system can therefore be quite large due to poorly-constrained assumptions made on the mass distribution \citep[for reviews, see e.g.][]{Jackson2015, DeGrijs2011, Schneider2013}.

\subsection{Differential Time Delays} \label{sec:timedelaydifferences}

\begin{figure}
    \centering
    \includegraphics[width=0.9\linewidth]{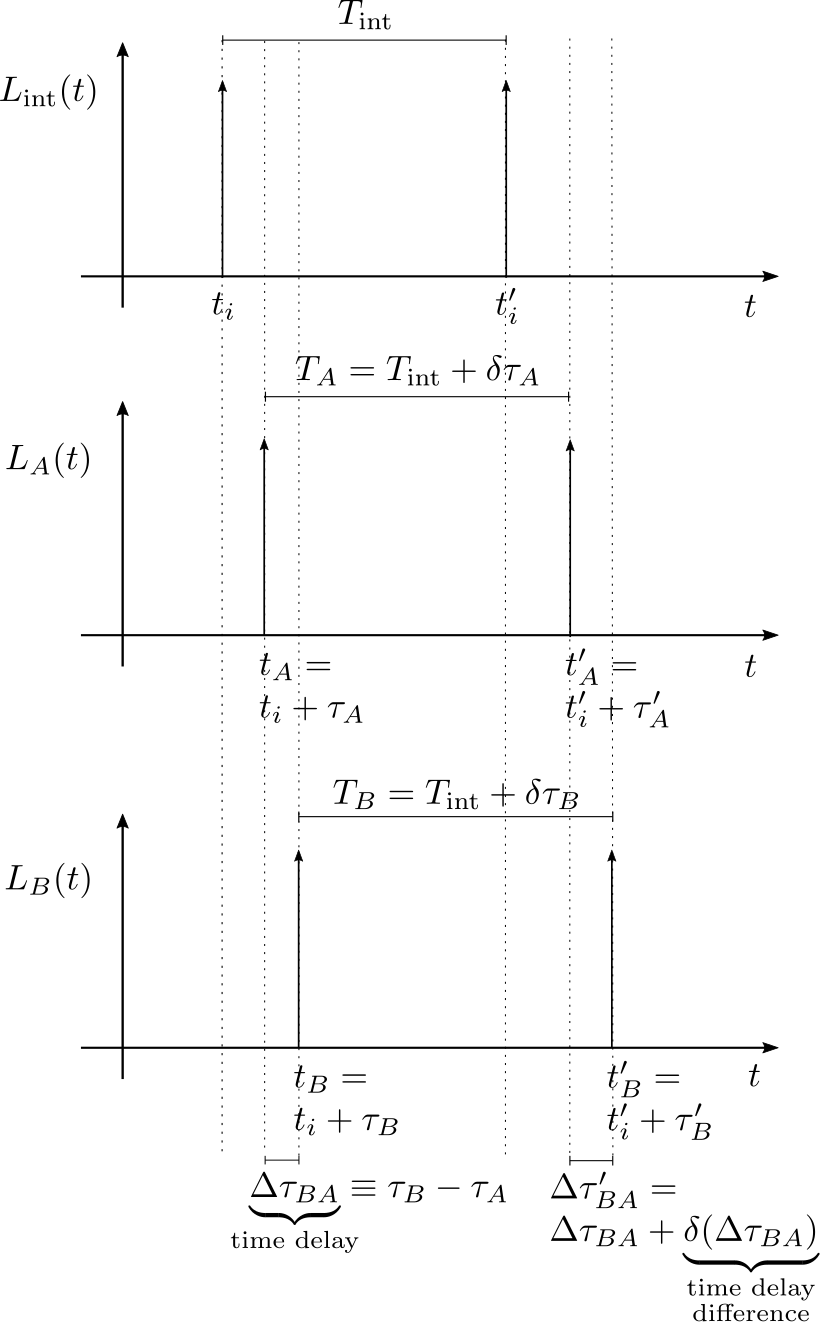}
    \caption{The luminosity of an unlensed source $L_{\mathrm{int}}(t)$ as seen by an observer, and of the corresponding lensed images A and B ($L_A (t)$ and $L_B (t)$ respectively), as a function of time. The signals corresponding to the unprimed times arise from a reference point in the source at $\vec{\beta}$, and the signals corresponding to primed times arise from a spatially perturbed location $\vec{\beta} + \delta \vec{\beta}$. The temporal intervals corresponding to the time delay and the time delay difference are marked.}
    \label{fig:timegraph}
\end{figure}

\begin{figure}
    \centering
    \includegraphics[width=\linewidth]{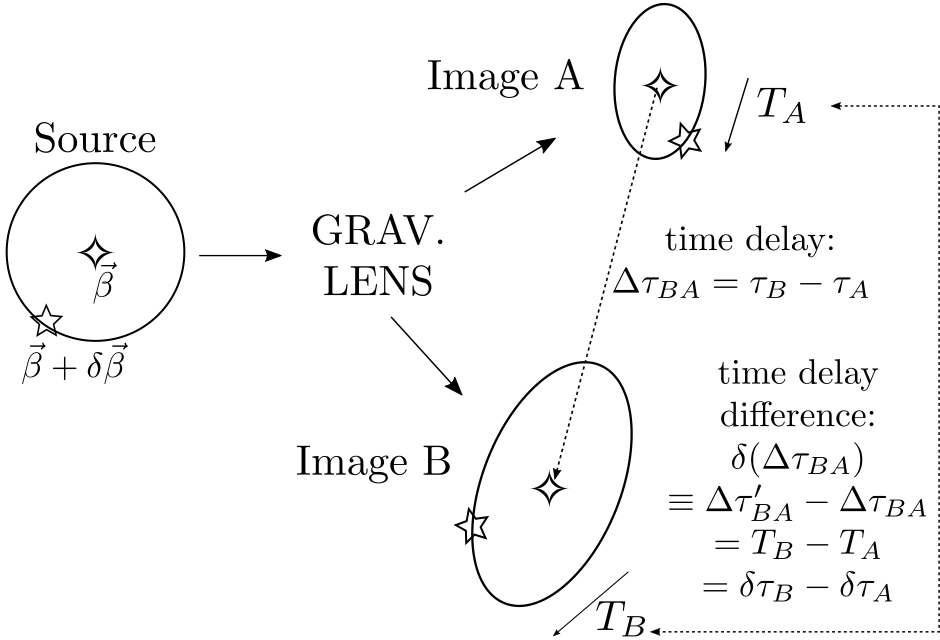}
    \caption{Schematic illustration of two flares, one at a reference point $\vec{\beta}$ in the source, and the other at a spatially perturbed location $\vec{\beta} + \delta \vec{\beta}$ in the source; and their corresponding signals in two lensed images \protect\citep[see][]{Yonehara1999}. The arrows corresponding to $T_A$ and $T_B$ are guides to show $T_X \equiv t_X' - t_X$, as opposed to $t_X - t_X'$.}
    \label{fig:spatialgraph}
\end{figure}

We present a method using differences in differential time delay measurements to probe cosmology. In the following, we show that this approach avoids entirely the systematic lens-modelling problem which usually dominates the error budget.

Let us model a quasar as a collection of point sources, surrounding the central engine. It is convenient to set the central engine as the reference point source at position and line-of-sight physical distance $(\vec{\beta}, l_s)$; we then consider any point source at $(\vec{\beta} + \delta \vec{\beta}, l_s + \delta l_s)$, spatially offset from the central engine. Should an observer measure a hypothetical unlensed signal from the spatially offset point at a time $t=t_i'$, then the lensed signal observed in a given image X would appear at time $t_X' = t_i' + \tau_X'$ and at a position $\vec{\theta}_X + \delta \vec{\theta}_X$. We use the following shorthand notation for the perturbed time delay:
\begin{align}
\begin{split}
    {\tau}'_X \equiv& \,\tau(\vec{\theta}_X + \delta \vec{\theta}_X , \vec{\beta} + \delta \vec{\beta}, l_s + \delta l_s)\\
    =& \, \frac{D_d F(l_s + \delta l_s)}{c} ( 1 + z_d) \Big( \frac{1}{2} \left((\vec{\theta}_X + \delta \vec{\theta}_X) - (\vec{\beta} + \delta \vec{\beta})\right)^2\\
    &- \Psi(\vec{\theta}_X + \delta \vec{\theta}_X) \Big).
\end{split}
\end{align}
where $F \equiv \frac{D_s}{D_{ds}}$.
We can similarly define a time delay between lensed images for the signal originating from the perturbed source,
\begin{equation}
    \Delta {\tau }'_{BA} \equiv {\tau }'_B - {\tau }'_A.
\end{equation}
The situation is illustrated schematically in Figures \ref{fig:timegraph} and \ref{fig:spatialgraph}. We have further defined $T_{\mathrm{int}} \equiv t_i' - t_i$ as the interval between the hypothetical unlensed signals from the perturbed and reference point sources. From Figure \ref{fig:timegraph} we can see that the interval $T_X$ between signals in a given image X is given by
\begin{equation}
    T_X \equiv t_X' - t_X = {t}'_i - t_i + {\tau }'_X - \tau_X = T_{\mathrm{int}} + \delta \tau_X \label{T_X}
\end{equation}
where we have defined the difference in time delay between the perturbed and reference points \textit{within image X}, the differential time delay, as
\begin{equation}
    \delta \tau_X \equiv {\tau }'_X - \tau_X.
\end{equation}
If the signals are concurrent, and at the same line-of-sight distance within the source, then $T_{\mathrm{int}} =0$ and so $T_X = \delta \tau_X$, i.e. the interval between signals in image X is the differential time delay within image X.

Computing $\delta \tau_X$ to first order in $\delta \vec{\beta}$, $\delta l_s$ and $\delta \vec{\theta}_X$, as shown by \citet{Tie2018}, gives:
\begin{align}
\begin{split}
    \delta \tau_X = &\frac{D}{c}(1+z_d) \Bigg((\vec{\beta} - \vec{\theta}_X)\cdot \delta \vec{\beta}\\ 
    &+ \frac{\delta l_s}{F} \frac{dF}{dl_s} \left(\frac{1}{2} ( \vec{\theta}_X - \vec{\beta})^2 - \Psi(\vec{\theta_X})\right)\Bigg) \qquad \text{to 1\textsuperscript{st} order}.
\end{split}
\end{align}
However, the second term corresponding to a displacement of the source in the line of sight of the observer is very small; we can therefore disregard $\delta l_s$ for \textit{lensing} time delays:
\begin{align}
    \delta \tau_X &= \frac{D}{c}(1+z_d)(\vec{\beta} - \vec{\theta}_X)\cdot \delta \vec{\beta} && \text{to 1\textsuperscript{st} order}. \label{DeltaTX}
\end{align}

We furthermore define the \textit{time delay difference} between images A and B as
\begin{align}
\begin{split}
    \delta (\Delta \tau_{BA}) &\equiv \Delta {\tau}'_{BA} - \Delta \tau_{BA}\\
    & = T_B - T_A = \delta \tau_B - \delta \tau_A.
\end{split} \label{timedelaydifferenceequationexact}
\end{align}
We see that it is the difference in the differential time delays of an image pair. Importantly, the time delay difference is independent of $\vec{\beta}$ and $\Psi(\vec{\theta}_X)$ to first order in $\delta \vec{\beta}$ and $\delta \vec{\theta}_X$ \citep{Yonehara1999,Goicoechea2002,Yonehara2003}:
\begin{align}
    \delta (\Delta \tau_{BA}) &= \frac{D}{c}(1+z_d) \left( \vec{\theta}_A - \vec{\theta}_B \right) \cdot \delta \vec{\beta} && \text{to 1\textsuperscript{st} order}. \label{timedelaydifferenceequation}
\end{align}
The time delay difference $\delta (\Delta \tau_{BA})$ is solely determined by cosmology through the lensing distance $D$, the geometry of the lensing configuration and the spatial separation within the source $\delta \vec{\beta}$ in the plane of the sky. The lens redshift and image positions can be measured directly, with the only remaining unknown being $\delta \vec{\beta}$. Remarkably, this expression removes much of the uncertainties associated with lens modelling in using time delay measurements to constrain cosmology. Time delay difference measurements hence appear to be an ideal avenue for testing cosmological models if $\delta \vec{\beta}$ can be determined: we turn to this in the next section.

\section{Broad Line Region Reverberation Mapping} \label{sec:reverbmapping}

The lensing time delay $\tau_X$ measures the time delay between a lensed light ray and its hypothetical unlensed counterpart from the same point source. As we wish to compare arrival times for signals from spatially separated sources, we must include an extra geometric delay arising from the difference in path lengths for the unlensed rays. We also need to know the relative emission times of the signals from the different source positions. The combination of these factors gives the time interval $T_{\mathrm{int}}$ between the signals in the previous section. Furthermore, given Equation \eqref{timedelaydifferenceequation} for the time delay difference, we wish to determine the displacement $\delta \vec{\beta}$ within the source. Although quasars remain spatially unresolved even with the best telescopes and at the lowest redshifts, we are still able to determine the quasar structure.

The spectra of quasars have characteristic broad emission lines corresponding to gas clouds, known as the Broad Line Region (BLR), surrounding the central emitting accretion disk at some distance. Photons travelling outwards from the central source are absorbed and re-emitted by the BLR gas. The broad emission lines therefore respond to variations in the continuum luminosity of the central source with a time delay determined by the BLR geometry.

Using this measured time delay to deduce the geometry of the BLR is an established technique known as reverberation mapping \citep{Blandford1982, Peterson1993, Shen2015, Dexter2019}. Since the BLR response time is set by the speed of light, the BLR reverberation sets an absolute distance scale, ideal for determining $\delta \vec{\beta}$. As the distance scale of the BLR is much greater than the accretion disk, the resultant differential time delay effects are substantially larger than those considered in previous literature \citep[e.g.][]{Yonehara1999, Goicoechea2002, Tie2018}.

For a particular BLR cloud located at a spatial position $\vec{r}$ relative to the central source, the associated time delay relative to the central source is determined by the reverberation mapping constraint equation:
\begin{equation}
\tau_{\mathrm{BLR}} \equiv (1+z_s) \tilde{\tau}_{\mathrm{BLR}} = \frac{(1+z_s)}{c} \left( |\vec{r}| - \vec{r} \cdot \hat{n}  \right) \label{tauBLR}
\end{equation}
where $\hat{n}$ is the unit vector towards the observer. The first term is the time taken for a photon to travel from the central source to the BLR cloud, i.e. the difference in emission times. The second term is the extra time for a photon to travel from a given BLR particle to the observer, as compared with a photon from the central source to the observer. The cosmological redshift factor $(1+z_s)$ corrects the time delay in the quasar rest frame, $\tilde{\tau}_{\mathrm{BLR}}$, for the time delay as measured in our observer frame, $\tau_{\mathrm{BLR}}$. The time interval between an unlensed signal from a point source in the BLR at $(\vec{\beta} + \delta \vec{\beta}, l_s + \delta l_s)$ compared with an unlensed signal from the central engine at $(\vec{\beta}, l_s)$ is therefore $T_{\mathrm{int}} = t_i' - t_i =\tau_{\mathrm{BLR}}$.

The BLR may be approximated by a flat geometry in Keplerian orbit around a central black hole, e.g. \citet{Grier2017}. As a toy example of reverberation mapping, we therefore consider an infinitesimally thin BLR ring, at the same redshift as the central source, facing the observer plane with a constant linear number density. The observed luminosity of the BLR, proportional to the number of responding particles $N$ at a given time, is
\begin{equation}
    L_{\mathrm{ring}}(\tau_{\mathrm{BLR}}) \propto \frac{dN}{d \tilde{\tau}_{\mathrm{BLR}}} \propto  \delta \left(\tilde{\tau}_{\mathrm{BLR}} - \frac{R}{c} \right),  \label{Psithinring}
\end{equation}
i.e. a Dirac delta function at a single time delay value $\tilde{\tau}_{\mathrm{BLR}} = \frac{R}{c}$, where $R$ is the radial distance of the thin ring from the central engine. We recognise intuitively this time delay as the light travel time between the central source and any cloud on the BLR ring, in accordance with the constraint Equation \eqref{tauBLR}. In this simplest form of reverberation mapping, as is often used for estimating black hole masses, BLR time delay measurements directly give an estimate for an average BLR radius \citep[e.g.][]{Bentz2015, Kaspi2017, Shapovalova2017}. This is in contrast with more sophisticated geometric and kinematic models of the BLR, \citep[e.g.][]{Sturm2018, Mangham2019}, which we will explore in future work.

\section{Gravitationally Lensed Broad Line Region Reverberation Mapping} \label{sec:lensedrm}

\begin{figure}
    \centering
    \includegraphics[width=\linewidth]{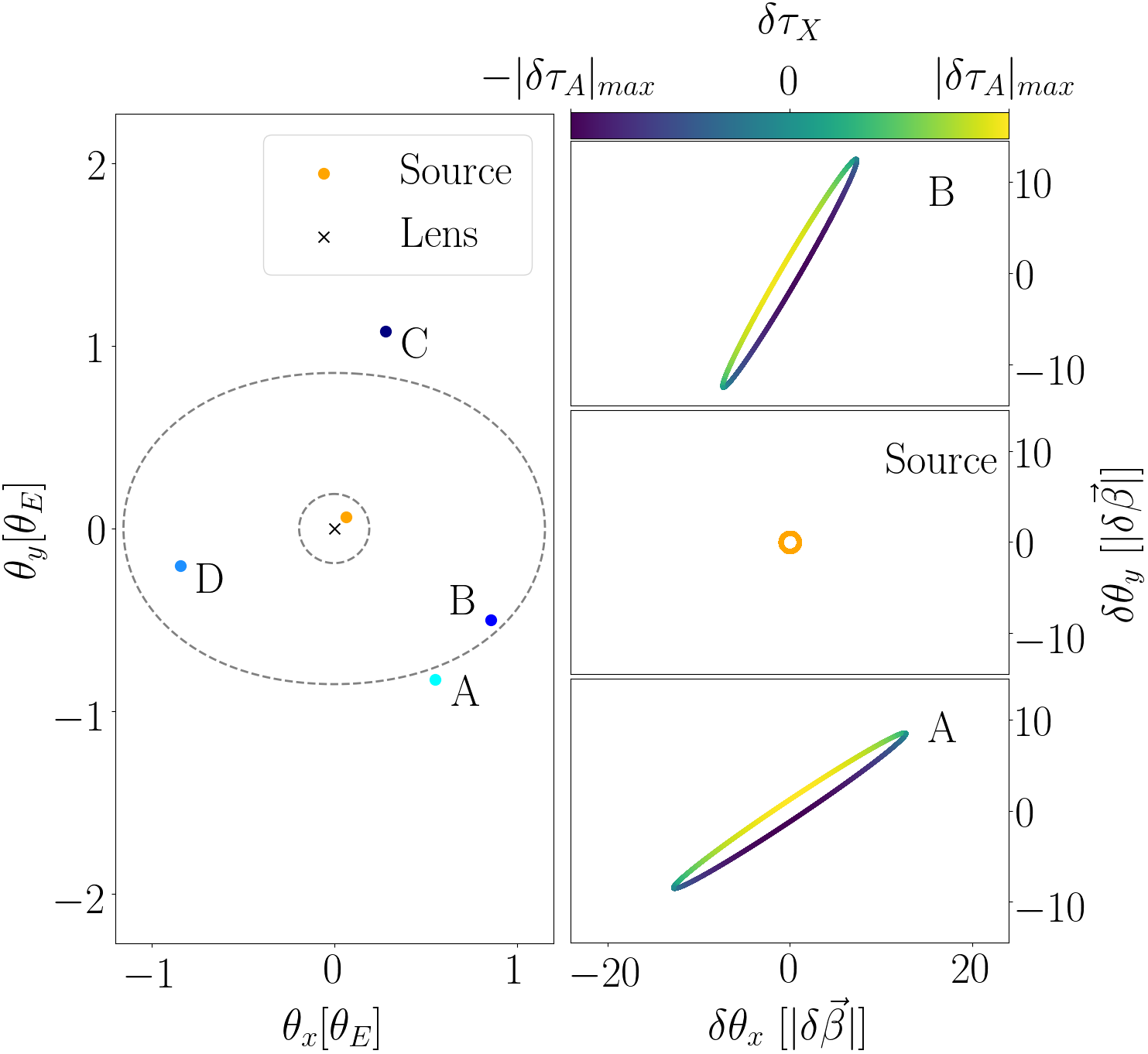}
    \caption{A typical configuration for a lensed quasar, showing the thin ring BLR structure of the quasar and its four images. The critical lines are marked by dashed grey lines. Image positions were calculated using a softened elliptical lens model with a core radius of $0.1 \theta_E$ and ellipticity of $0.1$, setting source at $\vec{\beta} = (0.065, 0.065) \theta_E$, where $\theta_E$ denotes the deflection scale of the lens. The differential time delays within images A and B (relative to the source centre) are shown; their respective maximum values are similar, since $|\vec{\beta} - \vec{\theta}_A| = 1.02 \theta_E \sim |\vec{\beta} - \vec{\theta}_B| = 0.97 \theta_E$.}\label{fig:lensingsetup}
\end{figure}

We continue to consider the toy model for the BLR outlined in the previous section. For a lensed quasar, we observe  multiple images where the BLR ring is distorted into ellipses (see Figure~\ref{fig:lensingsetup}). However, due to the effect of the differential time delay, only part of the ellipse emits the flux visible in a given moment of time. Each point within the lensed quasar images responds to the continuum emission in the source with a total time delay equal to the time delay from the BLR geometry $\tau_{\mathrm{BLR}}$ plus a time delay $\tau_X$ from lensing.

Recall that for an unlensed quasar, there is a Dirac delta signal from the central source at $t=t_i$ across all wavelengths (i.e. the continuum emission). This is followed by a Dirac delta signal from the entire BLR ring at $t= t_i' = t_i + \tau_{\mathrm{BLR}}$ in the broadened emission lines. The time that we see the continuum signal in image X is given by
\begin{equation}
    t_X = t_i + \tau_X. \label{timeXundashed}
\end{equation}
The time that we see a signal originating from a perturbed source position $(\vec{\beta} + \delta \beta, l_s + \delta l_s)$ in image X is similarly given by
\begin{equation}
    t_X' = t_i' + \tau_X' = t_i + \tau_{\mathrm{BLR}} + \tau_X + \delta \tau_X \label{timeXdashed}
\end{equation}
using $\tau_X' = \tau_X + \delta \tau_X$, as we note $\tau_X$ is constant in $\delta \vec{\beta}$ (as opposed to $\tau_X'$).

\begin{figure}
    \centering
    \includegraphics[width=\linewidth]{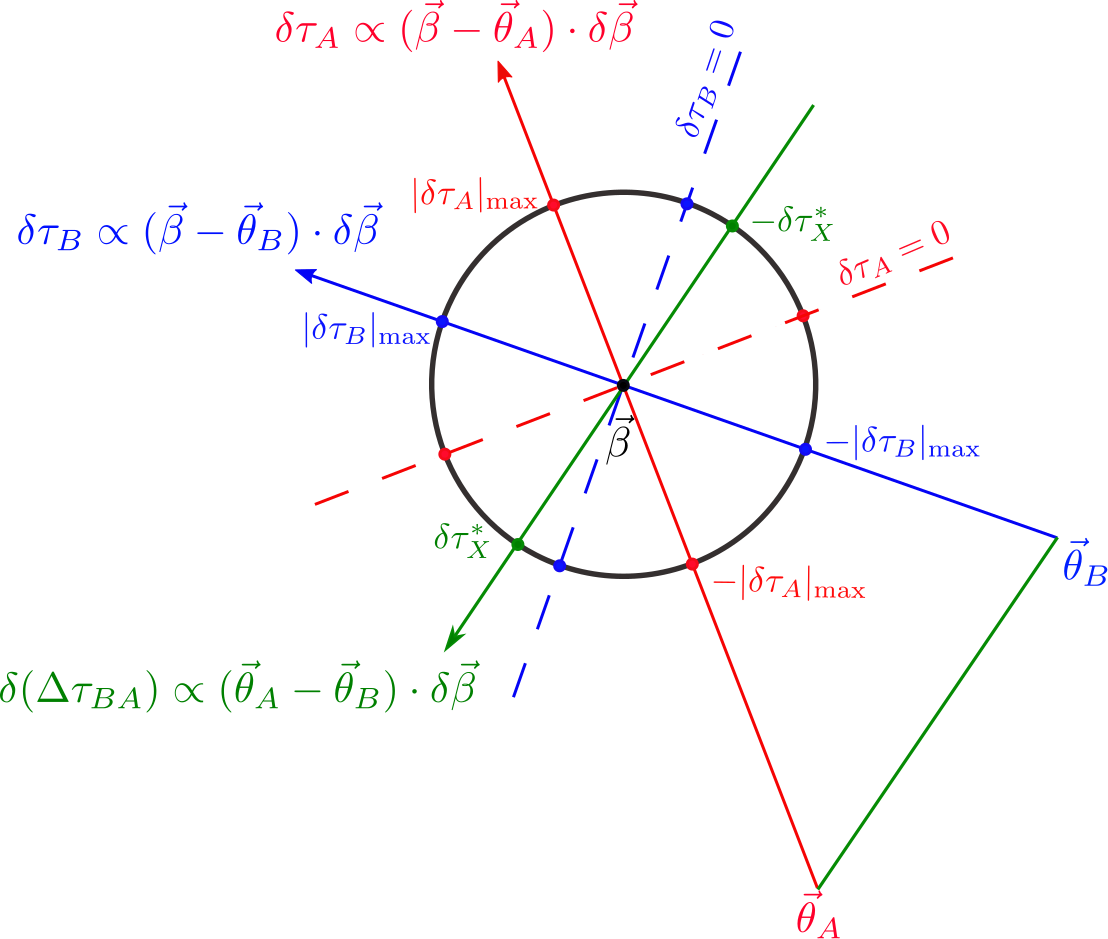}
    \caption{A representation of the source centred at $\vec{\beta}$ surrounded by the thin ring BLR (black circle), not to scale with the labelled image positions $\vec{\theta}_A$ and $\vec{\theta}_B$. Lines in the $\vec{\beta} - \vec{\theta}_A$ direction (red), $\vec{\beta} - \vec{\theta}_B$ direction (blue), and $\vec{\theta}_A - \vec{\theta}_B$ direction (green) through the source are labelled. The points in the source that lie furthest in the $\pm (\vec{\beta} - \vec{\theta}_X)$ direction map to points in image X with the maximal and minimal time delay respectively, $\delta \tau_X = \pm |\delta \tau_X |_\mathrm{max}$, compared to the source centre. The total differential time delay across image X is therefore $2|\delta \tau_X|_\mathrm{max}$. The linear dependence of $\delta \tau_X$ upon the component of the source position in the $\vec{\beta} - \vec{\theta}_X$ direction results in the arcsine distribution of the flux. Similarly, the points in the source that lie furthest in the $\pm (\vec{\theta}_A - \vec{\theta}_B)$ direction give the maximal and minimal time delay differences, $\pm |\delta(\Delta \tau_{BA})|_{\mathrm{max}}$, relative to the source centre. These points, when mapped to image X, have a differential time delay of $\pm \delta \tau_X^*$ respectively, as given by Equation \protect\eqref{deltatauasterisk}. The actual maximal measurable time delay difference is therefore $2|\delta(\Delta \tau_{BA})|_{\mathrm{max}} = 2 (\delta \tau_B^* - \delta \tau_A^*)$, arising from the difference in differential time delays from across the entire BLR diameter.}
    \label{fig:source}
\end{figure}

The variability in the BLR emission lines for a thin face-on ring geometry is a function of only $\delta \tau_X$ since $t_i$ and $\tau_X$ are fixed; and $\tau_{\mathrm{BLR}}$ is constant. We choose coordinates (for a particular image X) such that $\vec{\beta} - \vec{\theta}_X$ points in the $\hat{y}$ direction. From Equation \eqref{DeltaTX}, we have that $\delta \tau_X$ depends linearly on the $y$ component of $\delta \vec{\beta}$. As a result of this spatial dependence, only part of the BLR image emits the flux visible in a given moment of time; see Figures \ref{fig:source} and \ref{fig:lensingsetup}. 

To find the luminosity $L_X(t)$ in image X at time $t$ for a signal originating from an arbitrary location on the BLR ring, we need only find $L_X(\delta \tau_X)$. Let $\delta \vec{\beta} = (x,y)$ and using $\frac{dN}{dy} = \frac{2N}{2\pi|\delta \vec{\beta}|} \frac{|\delta \vec{\beta}|}{\sqrt{|\delta \vec{\beta}|^2 - y^2}}$ we have then that
\begin{align}
\begin{split}
    L_X(t) \propto \frac{dN}{d (\delta \tau_X)} &= \frac{1}{\frac{D}{c}(1+z_d) | \vec{\beta} - \vec{\theta}_X|} \frac{dN}{dy}\\
    &= \frac{N}{\pi \sqrt{|\delta \tau_X|_{\mathrm{max}}^2 - |\delta \tau_X|^2}} , \label{lensingthinringfunction}
\end{split}
\end{align}
where $|\delta \tau_X|_{\mathrm{max}} = \frac{D}{c}(1+z_d) | \vec{\beta} - \vec{\theta}_X| |\delta \vec{\beta}| $. \textit{Lensed BLR signals are therefore widened temporally into an arcsine distribution at each instant.}

\begin{figure}
    \centering
    \includegraphics[width=\linewidth]{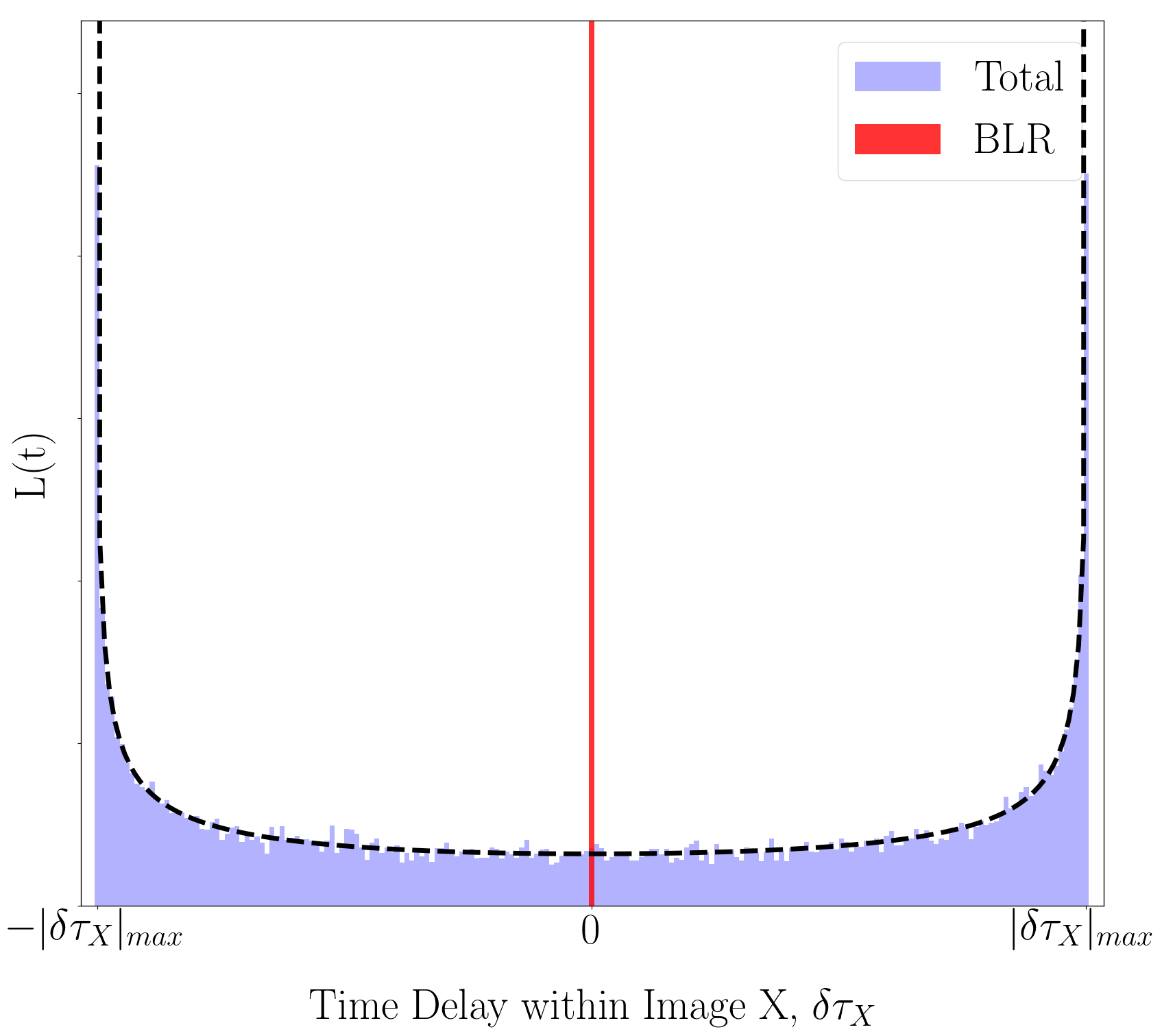}
    \caption{The effect of the differential time delay over an image on the reverberation mapping signal for the toy model considered in this paper. The unlensed view of the flux (red) is a Dirac delta function, as the entire BLR is seen to reverberate at the same time; this is distorted by differential lensing into the extended signal (blue). 
    The width of this signal corresponds to $2|\delta \tau_X|_{\mathrm{max}}$. The dashed line shows the prediction from Equation \protect\eqref{lensingthinringfunction}, an arcsine distribution.
    }
    \label{fig:luminosity_im0_sis}
\end{figure}

The situation is illustrated in Figure \ref{fig:luminosity_im0_sis}. In an image X, we see a Dirac delta function lensed continuum signal at time $t_X = t_i + \tau_X$, whereas the BLR signal in the emission lines is temporally widened into an arcsine distribution due to the spatially-dependent lensing time delay. The point on the ellipse that responds soonest does so at time $t_i + \tau_{\mathrm{BLR}} + \tau_X - |\delta \tau_X |_{\mathrm{max}}$ and the point on the ellipse that responds with the greatest time delay does so at $t_i + \tau_{\mathrm{BLR}} + \tau_X + |\delta \tau_X |_{\mathrm{max}}$. The arcsine distribution is therefore temporally centred at $\bar{t}_X' = t_i + \tau_{\mathrm{BLR}} + \tau_X$, with a width of $2|\delta \tau_X |_{\mathrm{max}}$.

We can measure $\tau_{\mathrm{BLR}}$, and hence $R$, by subtracting the time we see the lensed continuum signal in that image from the time at which the lensed BLR signal is centred: 
\begin{equation}
    |\bar{t}_X' - t_X |= |\tau_{\mathrm{BLR}}| = (1+z_s)\frac{R}{c}.
\end{equation}
We may thereby use reverberation mapping time delays to determine the source size
\begin{equation}
    |\delta \vec{\beta}|=\frac{R}{D_s} = \frac{c |\tau_{\mathrm{BLR}}|}{(1+z_s)D_s}
\end{equation}
of lensed quasars, and we obtain a separate measurement of $R$ for each image. Recalling Equation \eqref{timedelaydifferenceequation}, we have
\begin{equation}
   |\delta(\Delta \tau_{BA})|_{\mathrm{max}} = \frac{D_d}{D_{ds}} \frac{(1+z_d)}{(1+z_s)} |\vec{\theta}_A - \vec{\theta}_B | |\tau_{\mathrm{BLR}}|. \label{finaltimedelaydifferenceequation}
\end{equation}
The redshifts, the image separation $|\vec{\theta}_A - \vec{\theta}_B|$, and the BLR time delay $\tau_{\mathrm{BLR}}$ are measured quantities. Once the time delay difference $|\delta(\Delta \tau_{BA})|_{\mathrm{max}}$ is measured, the ratio of angular diameter distances $\frac{D_d}{D_{ds}}$ may be constrained.

We note that na{\"i}vely subtracting the widths $2|\delta \tau_X |_{\mathrm{max}}$ of lensed signals from two different images does not give a measurement of the time delay difference, including not of the maximum time delay difference: $|\delta(\Delta \tau_{BA})|_{\mathrm{max}} \neq |\delta \tau_B |_{\mathrm{max}} - |\delta \tau_A |_{\mathrm{max}}$. Rather, $|\delta (\Delta \tau_{BA})|_{\mathrm{max}} = \delta \tau_B^* - \delta \tau_A^*$ where $\delta \tau_X^*$ is given by
\begin{equation}
    \delta \tau_X^* = \frac{D}{c} (1+ z_d) \frac{(\vec{\beta} - \vec{\theta}_X) \cdot (\vec{\theta}_A - \vec{\theta}_B)}{|\vec{\theta}_A - \vec{\theta}_B|} |\delta \vec{\beta}| \label{deltatauasterisk}
\end{equation}
as illustrated in Figure \ref{fig:source}.

For more complicated BLR geometries we will need to distinguish the lensing effects from the response function of reverberation mapping. In general, the lensed luminosity function for an arbitrary geometry will be the convolution of the unlensed luminosity function for that geometry with the arcsine response corresponding to a face-on thin ring. Finding the appropriate timescale from the observed luminosity function corresponds to performing an inversion of this convolution, which may be a nontrivial task.

\section{A Measurement Method Using Lensed Quasar Spectra} \label{sec:spectra}

The spectra of lensed quasars contain more information than a measurement of the flux, as different spatial regions in the quasar are distinguishable through their velocities and hence the measured wavelengths. This additional information enables us to directly measure the differential time delays required to provide cosmological constraints. We leave the full kinematic signature, comprised of multiple spatial and velocity components at each instant, for a future contribution.

As a simplified example, we consider ``cloud 1'' at $\vec{r}_1$ and ``cloud 2'' at $\vec{r}_2$ within an arbitrary BLR geometry, possessing different velocities $\vec{v}_1$ and $\vec{v}_2$ relative to the central source. For a particular Doppler broadened emission line, we see an increased luminosity in the spectra of image A, at a wavelength corresponding to cloud 1, at time $t_{A,1}'$. Similarly, at time $t_{B,1}'$ we see a brightening in the spectra of image B at the same wavelength corresponding to cloud 1. At a time $t_{A,2}'$ we see an increased luminosity in the spectra of image A at a wavelength corresponding to cloud 2; this occurs in image B at a time $t_{B,2}'$.

Recalling $t_X'$ as given by Equation \eqref{timeXdashed}, we have
\begin{equation}
\begin{split}
    t_{X,1}' &= t_i + \tau_{\mathrm{BLR},1} + \tau_X + \delta \tau_{X,1}\\
    t_{X,2}' &= t_i + \tau_{\mathrm{BLR},2} + \tau_X + \delta \tau_{X,2}.
    \end{split}
\end{equation}
Since we are able to measure $t_{X,1}'$ and  $t_{X,2}'$ directly from the lensed spectra of each image, we are able to measure the differential time delays within each image
\begin{equation}
    t_{X,2}' - t_{X,1}' = (\tau_{\mathrm{BLR},2} - \tau_{\mathrm{BLR},1}) + (\delta \tau_{X,2} - \delta \tau_{X,1})
\end{equation}
and take the difference in the differential time delays between the two images
\begin{equation}
    (t_{B,2}' - t_{B,1}') - (t_{A,2}' - t_{A,1}') = (\delta \tau_{B,2} - \delta \tau_{B,1}) - (\delta \tau_{A,2} - \delta \tau_{A,1}).
\end{equation}

This is the time delay difference between clouds 1 and 2, and it will be measured to be a maximal value for a particular wavelength pair. These wavelengths correspond to clouds 1 and 2 located on opposite sides of the BLR in the direction of the image-image axis on the sky (see Figure~\ref{fig:source}), such that:
\begin{equation}
\begin{split}
    (\delta \tau_{B,2} - \delta \tau_{B,1}) - (\delta \tau_{A,2} - \delta \tau_{A,1}) & \underset{\mathrm{max}}{=} 2 \delta \tau_B^* - 2 \delta \tau_A^*\\
    &= 2 |\delta ( \Delta \tau_{BA} )|_{\mathrm{max}}
\end{split}
\end{equation}
where $\delta \tau_X^*$ is given by Equation \eqref{deltatauasterisk}.

Having obtained a measurement of the time delay difference, we have direct observational measurements of all components of Equation \eqref{finaltimedelaydifferenceequation} with the exception of the ratio of the angular diameter distances $\frac{D_d}{D_{ds}}$. This is a determination of cosmological distance ratios that is independent of astrophysical modelling.

\begin{figure}
    \centering
    \includegraphics[width=\linewidth]{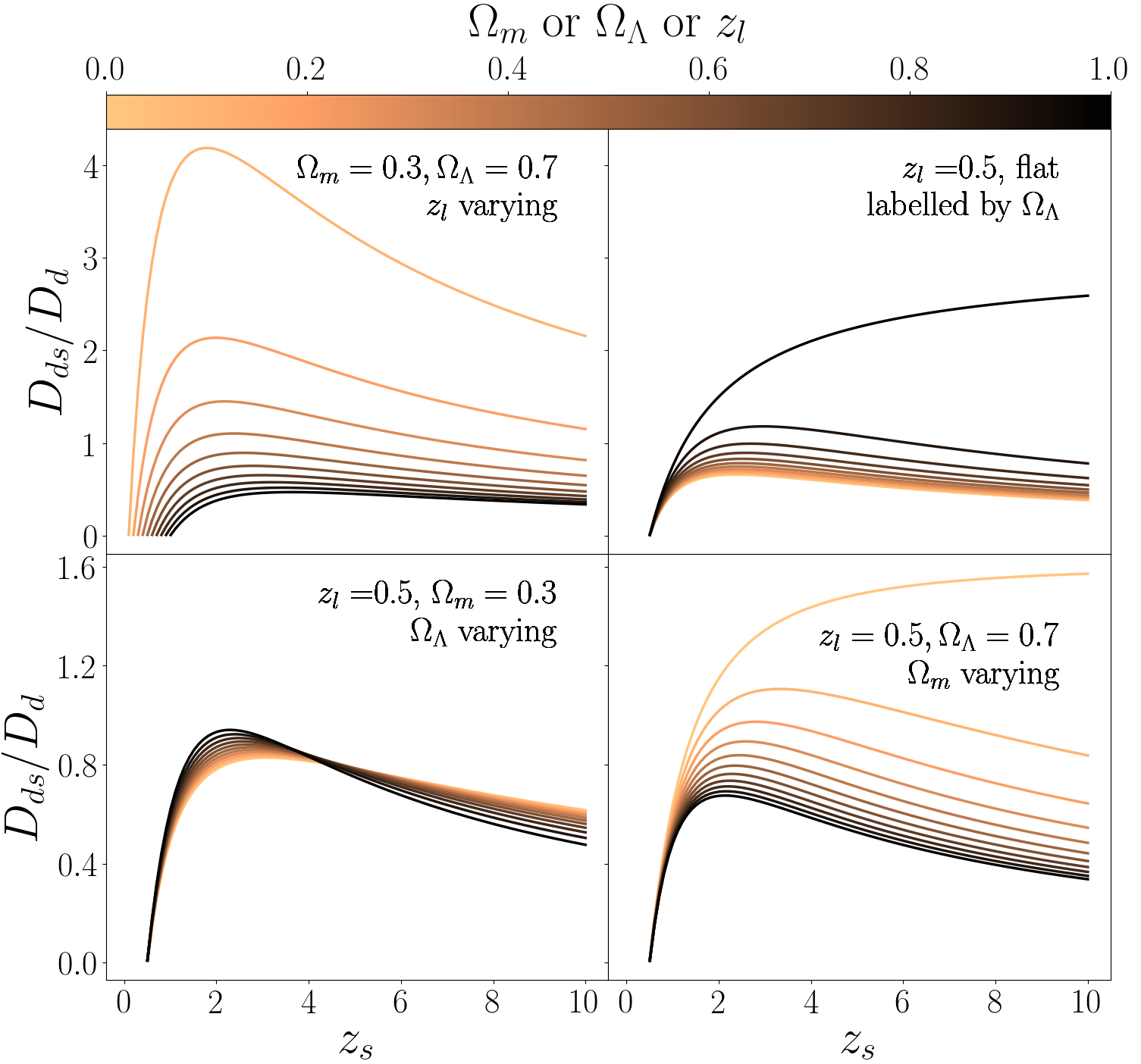}
    \caption{An illustration of the dependence of the ratio of the angular diameter distances $\frac{D_{ds}}{D_{d}}$ on the dark energy density and matter density parameters $\Omega_{\Lambda}$ and $\Omega_m$ assuming $\Lambda$CDM cosmologies. In top left plot we vary $z_l$ in equal intervals from $0.1$ to $1$ for a standard $\Lambda$CDM cosmology; in the top right plot we show the dependence of $\frac{D_{ds}}{D_{d}}$ on the dark energy density $\Omega_{\Lambda}$ (equivalently $\Omega_m$) under assumption of spatial flatness, $\Omega_m + \Omega_{\Lambda} =1$. In the bottom row, we allow curvature to vary from flat. For the top right plot and bottom row, lines are plotted in equal intervals from 0 to 1 in  $\Omega_M$ and $\Omega_{\Lambda}$.}
    \label{fig:paramdependence}
\end{figure}

The measurement of the differential time delay across images is a measure of a dimensionless ratio of angular diameter distances and thus insensitive to $H_0$; it is therefore complementary to probes of cosmology such as the standard time delay \citep[see][for a discussion on the dependencies of other distance ratios and the resulting degeneracies]{Linder2004, Linder2011}. In Figure~\ref{fig:paramdependence} we explore the sensitivity of this dimensionless ratio on various cosmological parameters, revealing that $\frac{D_{ds}}{D_{d}}$ is more sensitive to $\Omega_m$ than to $\Omega_{\Lambda}$, with a strong cosmological lever based upon the source redshift.

We note that by measuring both the conventional time delay between images $\Delta \tau_{BA} \propto \frac{D_d D_s}{D_{ds}}$ and the time delay difference $\delta( \Delta \tau_{BA} ) \propto \frac{D_d}{D_{ds}}$, we may obtain $\frac{\Delta \tau_{BA}}{\delta( \Delta \tau_{BA} )} \propto D_s$, i.e. the angular diameter distance to the source.

\section{Timescales \& Observability} \label{sec:timescales}
Typical BLR radii range from a few light days to a few hundred light days, as estimated from reverberation mapping \citep{Chelouche2014, Bentz2015, Kaspi2017, Shapovalova2017, Lira2018, Sturm2018}, corresponding to an angular extent on the sky on the order of \SIrange[range-units = single]{e-11}{e-10}{\radian}. For a $\Lambda$CDM cosmology considering typical lens redshifts of $0.3-1$ and source redshifts $1-4$ (and potentially up to $z\sim7$), we expect that $\frac{D}{c}(1+z_d)$ is on the order of \SIrange[range-units = single]{e17}{e18}{\second}. Recalling Equation \eqref{DeltaTX} and
Equation \eqref{timedelaydifferenceequation}, we have
\begin{align}
    |\delta \tau_X|_{\mathrm{max}} &\sim |(\vec{\beta} - \vec{\theta}_X)| \times (\num{e6} - \num{e8}) \si{\second}\\
    |\delta (\Delta \tau_{BA})|_{\mathrm{max}} &\sim  |( \vec{\theta}_A - \vec{\theta}_B)| \times (\num{e6} - \num{e8}) \si{\second} \label{orderofmagtimescales}
\end{align}
which sets the timescale for the differential time delay within the image and the timescale for the time delay difference between images. The minimum resolution at which we need to sample for the differential time delay is set by $2\delta \tau_X^*$ which will be some fraction of $2|\delta \tau_X|_{\mathrm{max}}$.

Considering cluster-scale lensing, with an angular deflection scale $\theta_E \sim 10 - 100''$, the largest image separations are twice the deflection scale, $\sim 20 - 200'' \sim$ \num{e-4}$ - $\SI{e-3}{\radian}. This gives a differential time delay within the images $\delta \tau$, as well as a time delay difference $\delta (\Delta \tau)$, on the order of hours to days. On a galaxy scale, the image separations are $\sim 2'' \sim$ \SI{e-5}{\radian}, giving a differential time delay and time delay difference on the order of minutes.

As an example, consider a high redshift quasar at $z_s = 3$ with a $200$ light day BLR radius, lensed by a mass at $z_l \sim 1$, with two images C and D separated by $\sim 1.7 \theta_E$ as illustrated in Figure~\ref{fig:lensingsetup}. For a galaxy scale lens, the differential time delay across the image is $2 |\delta \tau_X|_{\mathrm{max}} \sim  2 \delta \tau_X^* \sim$ \SI{12}{\min}, with a time delay $\Delta \tau_{DC}$ between images on the order of a month. For small clusters, we have $2|\delta \tau_X|_{\mathrm{max}} \sim 2\delta \tau_X^* \sim$ \SI{2}{\hour}; whilst the time delay $\Delta \tau_{DC}$ is on the order of a decade. For a very large cluster lens, the differential time delay within each image is $2|\delta \tau_X|_{\mathrm{max}} \sim  2 \delta \tau_X^* \sim$ \SI{20}{\hour}. However, the time delay $\Delta \tau_{DC}$ is on the order of hundreds to thousands of years, rendering measurements unattainable.

The most promising scenarios for measuring the standard time delays between images $\Delta \tau$ as well as differential time delays $\delta \tau$ are those for which the $\delta \tau$ timescale is still significant whilst reducing the $\Delta \tau$ timescale. In addition to using quasars with emission lines corresponding to clouds at larger radii, we may look for elliptical lenses where image separations can approach zero as the source crosses caustics. Such configurations can result in reasonable timescales for both differential time delays and time delays between images. For example, the quintuply cluster-lensed quasar SDSS J1004+4112 has a close image pair separated by only $3.8''$, such that the time delay between the two images is on the order of 40 days \citep{Fohlmeister2007, Fohlmeister2008}, and the same order of magnitude time delays between close image pairs occur in the sextuply cluster-lensed quasar SDSS J2222+2745 \citep{Dahle2015, Sharon2017}.

To demonstrate using the same example configuration, consider images A and B, with a much smaller separation $\sim 0.4 \theta_E$. For a galaxy scale lens, the differential time delay is $2|\delta \tau_X|_{\mathrm{max}} \sim$ \SI{20}{\min}, with $2 \delta \tau_A^* \sim$ \SI{4}{\min}, $2 \delta \tau_B^* \sim$ \SI{2}{\min}, and the time delay between images is $\Delta \tau_{BA} \sim$ \SI{10}{\hour}. For a small cluster, we have $2|\delta \tau_X|_{\mathrm{max}} \sim$ \SI{2}{\hour}, $2 \delta \tau_A^* \sim$ \SI{40}{\min}, $2 \delta \tau_B^* \sim$ \SI{15}{\min}, and $\Delta \tau_{BA}$ on the order of a month. On very large cluster scales we see a differential time delay on the order of $2|\delta \tau_X|_{\mathrm{max}} \sim$ \SI{20}{\hour}, with $2 \delta \tau_A^* \sim$ \SI{7}{\hour}, $2 \delta \tau_B^* \sim$ \SI{3}{\hour}; whereas the time delay is on the scale of a decade.

The measurement of these time delays within images and time delay differences between images are therefore reliant on the availability of light curves and spectra sampled with a frequency on the order of minutes. Long-term optical variability surveys such as The Optical Gravitational Lensing Experiment \citep[OGLE;][]{Udalski2015} and the Panoramic Survey Telescope and Rapid Response System \citep[Pan-STARRS;][]{Chambers2016} are ideal for measuring time delays in already-monitored lensed quasars, in addition to the dedicated COSMOGRAIL program \citep{Courbin2004}. Recent discoveries of lensed quasars using these variable sky surveys, such as \citet{Kostrzewa-Rutkowska2018} and in particular multiply imaged quasars such as \citet{Berghea2017} are very promising for time delay measurement endeavours. Spectroscopic data needed for time delay difference measurements is obtainable through the Time-Domain Spectroscopic Survey \citep[TDSS;][]{Ruan2016} for photometrically variable targets, including quasars, selected from e.g. Pan-STARRS and archival Sloan Digital Sky Survey imaging. 

The limiting factor of variable sky surveys is the relatively poor time sampling currently available. Quasars behind the Magellanic Clouds detected with OGLE are likely to remain amongst the most densely sampled (\num{1}$-$\SI{3}{\day}) long-term light curves available until the advent of the Large Synoptic Survey Telescope \citep[LSST;][]{Kozowski2013}. The main survey of the LSST \citep{Ivezic2008} will monitor the transient optical sky by tiling the sky with images of approximately ten-square-degrees, with two ``visits" (a pair of 15 second exposures per visit) on a given night, separated by 15-60 minutes \citep{lsst2009}. \citet{Oguri2010} discusses prospects for detecting strongly lensed time-variable sources in current and future time-domain optical imaging surveys; the number is estimated to be $\sim 3000$ for the LSST. The LSST project will furthermore include one hour of intensive observation per night of a set of ``Deep Drilling Fields" (DDFs) which will provide more frequent temporal sampling than the main survey. 50 consecutive 15-second exposures could be obtained in each of four filters in an hour, providing light curves of objects on hour-long timescales, which would be ideal for measuring the difference in time delay within images of lensed quasars; exactly how the LSST observations will be taken and the details of these intensive observations are not yet finalised \citep{lsst2017}.

\section{Conclusions and Future Work} \label{sec:conclusions}

Measurements of differential time delays across the gravitationally lensed images of reverberating BLR quasars offers a new and purely geometric test of cosmology. Such tests are highly relevant in an era when independent determinations of cosmological parameters are crucial to resolving tensions and bypassing systematic challenges associated with conventional methods. The method we have presented of measuring the time delay difference averts the degeneracies and difficulties inherent in lens modelling that typically affect time delay cosmography.

The time delay difference is determined by cosmology via a ratio of angular diameter distances, the image separation on the sky, and the spatial separation within the source. We have shown that gravitationally lensed reverberation-mapped quasars may be used as a means of constraining the source size. The BLR radius of the quasar may be measured, in the simplified case of our toy BLR model, directly from the difference in observed time between the lensed continuum signal and the centre of lensed BLR signal.

Furthermore, the spectra of the lensed quasar may be used to measure the time delay difference, as it carries additional information, the velocity and hence the spatial origin of the BLR cloud, via the wavelength of each signal. The result is that the ratio of angular diameter distances, $\frac{D_d}{D_{ds}}$, may be determined. This is a dimensionless ratio that depends most strongly on the matter density parameter, as well as the dark energy density parameter, and its determination will provide constraints on cosmology complementary to those from standard time delay measurements.

A thorough analysis of the impact of microlensing upon the signatures of the differential time delay will be conducted in a future paper. In particular, we expect that microlensing would impact observations of the continuum and line emissions differently, since the central source is smaller than the BLR. However, microlensing effects may be distinguished from the intrinsic quasar variability since variations from microlensing will be uncorrelated in different quasar images, whereas intrinsic variations appear in both images at different times. Furthermore, the chromatic effects of microlensing will be reduced by considering sources with larger radii (compared with the scale of the Einstein radius of typical microlenses in the source plane). Finally, we may choose lensing systems where the optical depth to microlensing is small, such as systems where images appear in the outer regions of the lensing galaxy.

We will also investigate in future work the sensitivity of the ratio $\frac{D_d}{D_{ds}}$ to cosmological parameters. We will include image configurations from realistic lenses and consider the measurement of time delay differences from highly-magnified images resulting from the source passing through a caustic. We will also consider realistic BLR models possessing a velocity function and include reverberation signals passing through geometric features in the BLR; and consider the effect of the amplitude of quasar variability on the relevant time scales.

\section*{Acknowledgements}
A.L.H.N. acknowledges funding from the Australian Government and The University of Sydney through a Research Training Program scholarship and a Hunstead Merit Award respectively.

%%%%%%%%%%%%%%%%%%%%%%%%%%%%%%%%%%%%%%%%%%%%%%%%%%



\bibliographystyle{mnras}
\bibliography{timedelayredshifts} % if your bibtex file is called example.bib

\begin{thebibliography}{}
\makeatletter
\relax
\def\mn@urlcharsother{\let\do\@makeother \do\$\do\&\do\#\do\^\do\_\do\%\do\~}
\def\mn@doi{\begingroup\mn@urlcharsother \@ifnextchar [ {\mn@doi@}
  {\mn@doi@[]}}
\def\mn@doi@[#1]#2{\def\@tempa{#1}\ifx\@tempa\@empty \href
  {http://dx.doi.org/#2} {doi:#2}\else \href {http://dx.doi.org/#2} {#1}\fi
  \endgroup}
\def\mn@eprint#1#2{\mn@eprint@#1:#2::\@nil}
\def\mn@eprint@arXiv#1{\href {http://arxiv.org/abs/#1} {{\tt arXiv:#1}}}
\def\mn@eprint@dblp#1{\href {http://dblp.uni-trier.de/rec/bibtex/#1.xml}
  {dblp:#1}}
\def\mn@eprint@#1:#2:#3:#4\@nil{\def\@tempa {#1}\def\@tempb {#2}\def\@tempc
  {#3}\ifx \@tempc \@empty \let \@tempc \@tempb \let \@tempb \@tempa \fi \ifx
  \@tempb \@empty \def\@tempb {arXiv}\fi \@ifundefined
  {mn@eprint@\@tempb}{\@tempb:\@tempc}{\expandafter \expandafter \csname
  mn@eprint@\@tempb\endcsname \expandafter{\@tempc}}}

\bibitem[\protect\citeauthoryear{Ade et~al.,}{Ade
  et~al.}{2016}]{PlanckCollaboration2016}
Ade P. A.~R.,  et~al., 2016, \mn@doi [Astron. Astrophys.]
  {10.1051/0004-6361/201525830}, 594, A13

\bibitem[\protect\citeauthoryear{Alam et~al.,}{Alam et~al.}{2017}]{Alam2017}
Alam S.,  et~al., 2017, \mn@doi [MNRAS] {10.1093/mnras/stx721}, 470, 2617

\bibitem[\protect\citeauthoryear{Bentz \& Katz}{Bentz \&
  Katz}{2015}]{Bentz2015}
Bentz M.~C.,  Katz S.,  2015, \mn@doi [PASP] {10.1086/679601}, 127, 67

\bibitem[\protect\citeauthoryear{Berghea, Nelson, Rusu, Keeton  \&
  Dudik}{Berghea et~al.}{2017}]{Berghea2017}
Berghea C.~T.,  Nelson G.~J.,  Rusu C.~E.,  Keeton C.~R.,   Dudik R.~P.,  2017,
  \mn@doi [ApJ] {10.3847/1538-4357/aa7aa6}, 844, 90

\bibitem[\protect\citeauthoryear{Birrer et~al.,}{Birrer
  et~al.}{2019}]{Birrer2019a}
Birrer S.,  et~al., 2019, \mn@doi [MNRAS] {10.1093/mnras/stz200}, 484, 4726

\bibitem[\protect\citeauthoryear{Blandford \& McKee}{Blandford \&
  McKee}{1982}]{Blandford1982}
Blandford R.~D.,  McKee C.~F.,  1982, \mn@doi [ApJ] {10.1086/159843}, 255, 419

\bibitem[\protect\citeauthoryear{Blandford \& Narayan}{Blandford \&
  Narayan}{1992}]{BlandfordNarayan1992}
Blandford R.~D.,  Narayan R.,  1992, \mn@doi [ARA{\&}A]
  {10.1146/annurev.astro.30.1.311}, 30, 311

\bibitem[\protect\citeauthoryear{Bonvin et~al.,}{Bonvin
  et~al.}{2017}]{Bonvin2017}
Bonvin V.,  et~al., 2017, \mn@doi [MNRAS] {10.1093/mnras/stw3006}, 465, 4914

\bibitem[\protect\citeauthoryear{Chambers et~al.,}{Chambers
  et~al.}{2016}]{Chambers2016}
Chambers K.~C.,  et~al., 2016, arXiv e-prints, (arXiv:1612.05560)

\bibitem[\protect\citeauthoryear{Chelouche, Shemmer, Cotlier, Barth  \&
  Rafter}{Chelouche et~al.}{2014}]{Chelouche2014}
Chelouche D.,  Shemmer O.,  Cotlier G.~I.,  Barth A.~J.,   Rafter S.~E.,  2014,
  \mn@doi [ApJ] {10.1088/0004-637X/785/2/140}, 785, 140

\bibitem[\protect\citeauthoryear{Courbin, Eigenbrod, Vuissoz, Meylan  \&
  Magain}{Courbin et~al.}{2004}]{Courbin2004}
Courbin F.,  Eigenbrod A.,  Vuissoz C.,  Meylan G.,   Magain P.,  2004, \mn@doi
  [Proc. Int. Astron. Union] {10.1017/S1743921305002097}, 2004, 297

\bibitem[\protect\citeauthoryear{Courbin et~al.,}{Courbin
  et~al.}{2018}]{Courbin2018}
Courbin F.,  et~al., 2018, \mn@doi [Astron. Astrophys.]
  {10.1051/0004-6361/201731461}, 609, A71

\bibitem[\protect\citeauthoryear{Dahle, Gladders, Sharon, Bayliss  \&
  Rigby}{Dahle et~al.}{2015}]{Dahle2015}
Dahle H.,  Gladders M.~D.,  Sharon K.,  Bayliss M.~B.,   Rigby J.~R.,  2015,
  \mn@doi [ApJ] {10.1088/0004-637X/813/1/67}, 813, 67

\bibitem[\protect\citeauthoryear{{De Grijs}}{{De Grijs}}{2011}]{DeGrijs2011}
{De Grijs} R.,  2011, {An Introduction to Distance Measurement in Astronomy}.
John Wiley {\&} Sons, Ltd, Chichester, UK, \mn@doi{10.1002/9781119978176}, \url
  {http://doi.wiley.com/10.1002/9781119978176}

\bibitem[\protect\citeauthoryear{Dexter et~al.,}{Dexter
  et~al.}{2019}]{Dexter2019}
Dexter J.,  et~al., 2019, arXiv e-prints, (arXiv:1906.10138v1)

\bibitem[\protect\citeauthoryear{{Di Valentino}, Melchiorri  \& Silk}{{Di
  Valentino} et~al.}{2016}]{DiValentino2016}
{Di Valentino} E.,  Melchiorri A.,   Silk J.,  2016, \mn@doi [Phys. Lett. B]
  {10.1016/j.physletb.2016.08.043}, 761, 242

\bibitem[\protect\citeauthoryear{Falco, Gorenstein  \& Shapiro}{Falco
  et~al.}{1985}]{Falco1985}
Falco E.~E.,  Gorenstein M.~V.,   Shapiro I.~I.,  1985, \mn@doi [ApJ]
  {10.1086/184422}, 289, L1

\bibitem[\protect\citeauthoryear{Fohlmeister et~al.,}{Fohlmeister
  et~al.}{2007}]{Fohlmeister2007}
Fohlmeister J.,  et~al., 2007, \mn@doi [ApJ] {10.1086/518018}, 662, 62

\bibitem[\protect\citeauthoryear{Fohlmeister, Kochanek, Falco, Morgan  \&
  Wambsganss}{Fohlmeister et~al.}{2008}]{Fohlmeister2008}
Fohlmeister J.,  Kochanek C.~S.,  Falco E.~E.,  Morgan C.~W.,   Wambsganss J.,
  2008, \mn@doi [ApJ] {10.1086/528789}, 676, 761

\bibitem[\protect\citeauthoryear{Goicoechea}{Goicoechea}{2002}]{Goicoechea2002}
Goicoechea L.~J.,  2002, \mn@doi [MNRAS] {10.1046/j.1365-8711.2002.05574.x},
  334, 905

\bibitem[\protect\citeauthoryear{Grier, Pancoast, Barth, Fausnaugh, Brewer,
  Treu  \& Peterson}{Grier et~al.}{2017}]{Grier2017}
Grier C.~J.,  Pancoast A.,  Barth A.~J.,  Fausnaugh M.~M.,  Brewer B.~J.,  Treu
  T.,   Peterson B.~M.,  2017, \mn@doi [ApJ] {10.3847/1538-4357/aa901b}, 849,
  146

\bibitem[\protect\citeauthoryear{Ili{\'{c}} et~al.,}{Ili{\'{c}}
  et~al.}{2017}]{Shapovalova2017}
Ili{\'{c}} D.,  et~al., 2017, \mn@doi [Front. Astron. Sp. Sci.]
  {10.3389/fspas.2017.00012}, 4, 1

\bibitem[\protect\citeauthoryear{Ivezi{\'{c}} et~al.,}{Ivezi{\'{c}}
  et~al.}{2019}]{Ivezic2008}
Ivezi{\'{c}} {\v{Z}}.,  et~al., 2019, \mn@doi [ApJ] {10.3847/1538-4357/ab042c},
  873, 111

\bibitem[\protect\citeauthoryear{Jackson}{Jackson}{2015}]{Jackson2015}
Jackson N.,  2015, \mn@doi [Living Rev. Relativ.] {10.1007/lrr-2015-2}, 18, 2

\bibitem[\protect\citeauthoryear{Kaspi, Brandt, Maoz, Netzer, Schneider  \&
  Shemmer}{Kaspi et~al.}{2017}]{Kaspi2017}
Kaspi S.,  Brandt W.~N.,  Maoz D.,  Netzer H.,  Schneider D.~P.,   Shemmer O.,
  2017, \mn@doi [Front. Astron. Sp. Sci.] {10.3389/fspas.2017.00031}, 4, 1

\bibitem[\protect\citeauthoryear{Kostrzewa-Rutkowska
  et~al.,}{Kostrzewa-Rutkowska et~al.}{2018}]{Kostrzewa-Rutkowska2018}
Kostrzewa-Rutkowska Z.,  et~al., 2018, \mn@doi [MNRAS] {10.1093/mnras/sty259},
  476, 663

\bibitem[\protect\citeauthoryear{Koz{\l}owski et~al.,}{Koz{\l}owski
  et~al.}{2013}]{Kozowski2013}
Koz{\l}owski S.,  et~al., 2013, \mn@doi [ApJ] {10.1088/0004-637X/775/2/92}, 775

\bibitem[\protect\citeauthoryear{{LSST Science Collaboration} et~al.,}{{LSST
  Science Collaboration} et~al.}{2009}]{lsst2009}
{LSST Science Collaboration} et~al., 2009, arXiv e-prints, (arXiv:0912.0201),
  p.~1

\bibitem[\protect\citeauthoryear{{LSST Science Collaboration} et~al.,}{{LSST
  Science Collaboration} et~al.}{2017}]{lsst2017}
{LSST Science Collaboration} et~al., 2017, \mn@doi [arXiv e-prints,
  (arXiv:1708.04058)] {10.5281/zenodo.842713}, pp 1--312

\bibitem[\protect\citeauthoryear{Linder}{Linder}{2004}]{Linder2004}
Linder E.~V.,  2004, \mn@doi [Phys. Rev. D] {10.1103/PhysRevD.70.043534}, 70,
  043534

\bibitem[\protect\citeauthoryear{Linder}{Linder}{2011}]{Linder2011}
Linder E.~V.,  2011, \mn@doi [Phys. Rev. D] {10.1103/PhysRevD.84.123529}, 84,
  123529

\bibitem[\protect\citeauthoryear{Lira, Botti, Kaspi  \& Netzer}{Lira
  et~al.}{2018}]{Lira2018}
Lira P.,  Botti I.,  Kaspi S.,   Netzer H.,  2018, \mn@doi [Front. Astron. Sp.
  Sci.] {10.3389/fspas.2017.00071}, 4, 1

\bibitem[\protect\citeauthoryear{Mangham et~al.,}{Mangham
  et~al.}{2019}]{Mangham2019}
Mangham S.~W.,  et~al., 2019, \mn@doi [MNRAS] {10.1093/mnras/stz1713}

\bibitem[\protect\citeauthoryear{Oguri \& Marshall}{Oguri \&
  Marshall}{2010}]{Oguri2010}
Oguri M.,  Marshall P.~J.,  2010, \mn@doi [MNRAS]
  {10.1111/j.1365-2966.2010.16639.x}, 405, 2579

\bibitem[\protect\citeauthoryear{Peterson}{Peterson}{1993}]{Peterson1993}
Peterson B.~M.,  1993, \mn@doi [PASP] {10.1086/133140}, 105, 247

\bibitem[\protect\citeauthoryear{Refsdal}{Refsdal}{1964}]{Refsdal1964}
Refsdal S.,  1964, \mn@doi [MNRAS] {10.1093/mnras/128.4.307}, 128, 307

\bibitem[\protect\citeauthoryear{Riess et~al.,}{Riess et~al.}{2018}]{Riess2018}
Riess A.~G.,  et~al., 2018, \mn@doi [ApJ] {10.3847/1538-4357/aac82e}, 861, 126

\bibitem[\protect\citeauthoryear{Riess, Casertano, Yuan, Macri  \&
  Scolnic}{Riess et~al.}{2019}]{Riess2019}
Riess A.~G.,  Casertano S.,  Yuan W.,  Macri L.~M.,   Scolnic D.,  2019,
  \mn@doi [ApJ] {10.3847/1538-4357/ab1422}, 876, 85

\bibitem[\protect\citeauthoryear{Ruan et~al.,}{Ruan et~al.}{2016}]{Ruan2016}
Ruan J.~J.,  et~al., 2016, \mn@doi [ApJ] {10.3847/0004-637X/825/2/137}, 825,
  137

\bibitem[\protect\citeauthoryear{Saha}{Saha}{2000}]{Saha2000}
Saha P.,  2000, \mn@doi [AJ] {10.1086/301581}, 120, 1654

\bibitem[\protect\citeauthoryear{Schneider \& Sluse}{Schneider \&
  Sluse}{2013}]{Schneider2013}
Schneider P.,  Sluse D.,  2013, \mn@doi [Astron. Astrophys.]
  {10.1051/0004-6361/201321882}, 559, A37

\bibitem[\protect\citeauthoryear{Schneider, Ehlers  \& Falco}{Schneider
  et~al.}{1992}]{Schneider1992}
Schneider P.,  Ehlers J.,   Falco E.~E.,  1992, {Gravitational Lenses}.
Astronomy and Astrophysics Library, Springer Berlin Heidelberg, Berlin,
  Heidelberg, \mn@doi{10.1007/978-3-662-03758-4}, \url
  {http://link.springer.com/10.1007/978-3-662-03758-4}

\bibitem[\protect\citeauthoryear{Sharon et~al.,}{Sharon
  et~al.}{2017}]{Sharon2017}
Sharon K.,  et~al., 2017, \mn@doi [ApJ] {10.3847/1538-4357/835/1/5}, 835, 5

\bibitem[\protect\citeauthoryear{Shen et~al.,}{Shen et~al.}{2014}]{Shen2015}
Shen Y.,  et~al., 2014, \mn@doi [ApJS] {10.1088/0067-0049/216/1/4}, 216, 4

\bibitem[\protect\citeauthoryear{Sturm et~al.,}{Sturm et~al.}{2018}]{Sturm2018}
Sturm E.,  et~al., 2018, \mn@doi [Nature] {10.1038/s41586-018-0731-9}, 563, 657

\bibitem[\protect\citeauthoryear{Suyu et~al.,}{Suyu et~al.}{2014}]{Suyu2014}
Suyu S.~H.,  et~al., 2014, \mn@doi [ApJ] {10.1088/2041-8205/788/2/L35}, 788,
  L35

\bibitem[\protect\citeauthoryear{Suyu et~al.,}{Suyu et~al.}{2017}]{Suyu2017}
Suyu S.~H.,  et~al., 2017, \mn@doi [MNRAS] {10.1093/mnras/stx483}, 468, 2590

\bibitem[\protect\citeauthoryear{Tewes et~al.,}{Tewes et~al.}{2013}]{Tewes2013}
Tewes M.,  et~al., 2013, \mn@doi [Astron. Astrophys.]
  {10.1051/0004-6361/201220352}, 556, A22

\bibitem[\protect\citeauthoryear{Tie \& Kochanek}{Tie \&
  Kochanek}{2018}]{Tie2018}
Tie S.~S.,  Kochanek C.~S.,  2018, \mn@doi [MNRAS] {10.1093/mnras/stx2348},
  473, 80

\bibitem[\protect\citeauthoryear{Treu et~al.,}{Treu et~al.}{2013}]{Treu2013}
Treu T.,  et~al., 2013, arXiv e-prints, (arXiv:1306.1272)

\bibitem[\protect\citeauthoryear{Udalski, Szyma{\'{n}}ski  \&
  Szyma{\'{n}}ski}{Udalski et~al.}{2015}]{Udalski2015}
Udalski A.,  Szyma{\'{n}}ski M.~K.,   Szyma{\'{n}}ski G.,  2015, Acta Astron.,
  65, 1

\bibitem[\protect\citeauthoryear{Yonehara}{Yonehara}{1999}]{Yonehara1999}
Yonehara A.,  1999, \mn@doi [ApJ] {10.1086/312107}, 519, L31

\bibitem[\protect\citeauthoryear{Yonehara, Mineshige, Takei, Chartas  \&
  Turner}{Yonehara et~al.}{2003}]{Yonehara2003}
Yonehara A.,  Mineshige S.,  Takei Y.,  Chartas G.,   Turner E.~L.,  2003,
  \mn@doi [ApJ] {10.1086/376936}, 594, 107

\makeatother
\end{thebibliography}

%%%%%%%%%%%%%%%%%%%%%%%%%%%%%%%%%%%%%%%%%%%%%%%%%%

%%%%%%%%%%%%%%%%% APPENDICES %%%%%%%%%%%%%%%%%%%%%

\appendix

%%%%%%%%%%%%%%%%%%%%%%%%%%%%%%%%%%%%%%%%%%%%%%%%%%

% Don't change these lines
\bsp	% typesetting comment
\label{lastpage}
\end{document}